# Efficient Spatial Redistribution of Quantum Dot Spontaneous Emission from 2D Photonic Crystals

M. Kaniber, A. Kress, M. Bichler, R. Meyer,
M.-C. Amann and J. J. Finley

*Walter Schottky Institut and Physik Department, Technische Universität München,
Am Coulombwall 3, D-85748 Garching, Germany*

## Abstract

We investigate the modification of the spontaneous emission dynamics and external quantum efficiency for self-assembled InGaAs quantum dots coupled to extended and localised photonic states in GaAs 2D-photonic crystals. The 2D-photonic bandgap is shown to give rise to a 5-10 times enhancement of the external quantum efficiency whilst the spontaneous emission rate is simultaneously reduced by a comparable factor. Our findings are quantitatively explained by a modal redistribution of spontaneous emission due to the modified local density of photonic states. The results suggest that quantum dots embedded within 2D-photonic crystals are suitable for practical single photon sources with high external efficiency.

The ability to manipulate the spontaneous emission rate in solids and funnel energy into specific optical modes has strong potential for many applications across the fields of integrated photonics and quantum optics. For example, in the development of efficient light emitters[1,2,3] or low threshold lasers[4,5], photons which are not extracted from the device constitute a major source of loss. In these respects, a central concept is the use of photonic bandgap (PBG) materials to modify the total spontaneous emission (SE) rate and redistribute emission into specific optical modes.[1,6,7,8] The need for the highest possible external quantum efficiency ($\xi$) is, perhaps, most critical in the development of solid-state single photon sources.[9,10,11] A number of groups have already demonstrated efficient devices based on single, self-assembled quantum dots (QDs) coupled to high finesse optical nano-cavities.[9,10,12] Whilst such approaches are rather successful with $\xi$ up to a few percent, they call for significant technological effort since QDs must be both spectrally and spatially coupled to the nano-cavity mode.[13,14] However, for many applications in quantum cryptography the high degree of quantum indistinguishability afforded by such approaches is of secondary importance compared to the need for photon extraction efficiency. Practical sources that are easy to fabricate would offer significant advantages.

In this letter, we report the use of two-dimensional PBG materials to control the directionality and extraction efficiency of SE from InGaAs QDs. We demonstrate that the PBG alone results in a ~5-10 fold *increase* of the photon extraction efficiency, whilst the SE-rate is simultaneously *reduced* by more than an order of magnitude. The enhancement of $\xi$ due to the PBG is shown to be comparable to dots coupled to high-Q optical nanocavities indicating that simple 2D-PC structures containing QDs are suitable for the realisation of high efficiency single photon sources.

The samples investigated consisted of a 180nm thick GaAs slab-waveguide grown by molecular beam epitaxy (MBE) and clad by air on the upper surface and a 1μm thick sacrificial $Al_{0.8}Ga_{0.2}As$ layer on the substrate side. A single layer of InGaAs QDs was embedded 90nm below the surface in the centre of the GaAs waveguide. Two-dimensional photonic crystals (2D-PCs) were then fabricated using the following process steps: Firstly, a 80nm thick $Si_3N_4$ hard mask was deposited on the sample surface. A hexagonal lattice of circular holes was then patterned into this layer using electron beam lithography and $CF_4$ reactive ion etching, before this pattern was transferred through the waveguide core using a second $Cl_2$/Ar etching step. This produced a hexagonal array of air holes, each with radius *r*, forming the PC nanostructure. Finally, wet etching was used to selectively remove the $Al_{0.8}Ga_{0.2}As$ layer and form a freestanding GaAs membrane. This structure provides optical confinement along the MBE growth (*z*) direction due to total internal reflection at the GaAs-air interface and a modified photonic density of states in the in-plane (*x,y*) directions due to the PBG for TE-like guided modes in the slab.[15] A single reduced symmetry missing hole defect was defined in the centre of each photonic crystal field (fig 1a-inset). This defect nanocavity supports two high-Q resonator modes (*M1* and *M2*, $Q=10^3-10^4$) with a low mode volume $V_{mode} \sim 0.5(\lambda_{cavity}/n_{eff})^3$.[16]

Fig 1a (inset) depicts a top view SEM image of the sample investigated showing the different regions studied in this letter: **(i)** the defect nano-cavity, **(ii)** the 2D-PC, **(iii)** the unpatterned GaAs membrane and **(iv)** the unprocessed GaAs-AlGaAs waveguide. The lattice constant of the 2D-PC was chosen to be *a=280*nm and the air hole radius was systematically varied in the range *0.28 < r/a < 0.32* in order to tune the PBG such that it overlaps with the inter-band emission from the InGaAs QDs. We performed 3D calculations of the photonic bandstructure for the range of structural parameters investigated. The results of these calculations are summarised in fig 1a, showing clearly the air and dielectric continuum band edges and the 2D-PBG extending from *~915* to *1079*nm for *r/a=0.3*. The cavity modes *M1* and *M2* shift parallel to the band edges, within the PBG.[17]

We performed spatially resolved μ-photoluminescence (μ-PL) measurements using a confocal system that provides a spatial resolution of ~700nm. The sample is held inside a He-flow cryostat at T=10K and is optically excited using ~50ps duration pulses at 658nm from a diode laser. The resulting luminescence was spectrally dispersed by a 0.55m monochromator and detected using a Si-based CCD camera or avalanche photodiodes for time resolved measurements. This system provides a temporal resolution of ~150ps after Deconvolution with the instrument response function.

We begin by presenting CW μ-PL measurements recorded from position (**i**) in the vicinity of the nanocavity. Under strong CW optical pumping the QDs emit preferentially into the cavity modes due to the Purcell enhancement of the emission rate for dots *spectrally* and *spatially* coupled to the resonator.[18] Fig 1b shows a μ-PL spectrum showing *M1* and *M2* at $\lambda_{M1}$=1000±0.2 nm ($Q_{M1}$~6500) and $\lambda_{M2}$=1054.5±0.2 nm ($Q_{M2}$~5000), respectively.[19] From plan view SEM microscopy we measured *r/a*=0.30±0.01 on the body of the 2D-PC, in excellent agreement with the observed cavity mode wavelengths (fig 1a).

We continue to present time-resolved measurements recorded from the structure for different spatial positions (**i-iv**) and detection wavelengths ($\lambda_{det}$) both inside and outside the PBG. For reference, we began by measuring a series of time-resolved traces within a Δλ=1nm bandwidth from QDs in the unpatterned bulk material. A typical example of these measurements is presented in the upper panel of fig 2a for $\lambda_{det}$=1000nm. A clear mono-exponential decay is observed from which we obtain the rate $\tau_0^{-1}$ ~1.31±0.02 ns$^{-1}$, typical for such self-assembled QDs. The spectral dependence of $\tau_0^{-1}$ is plotted in the upper panel of fig 2b (filled circles) showing that it remains approximately constant over the whole wavelength range studied. Measurements on the unpatterned membrane (not presented here) produced comparable results, showing that the waveguide has little influence on the SE dynamics.

In strong contrast, moving from the unpatterned membrane onto the body of the PC was found to result in a drastic change in the dynamics. The middle panel of fig 2a shows a decay transient recorded at $\lambda_{det}$=1000nm for comparison with the data recorded from the unpatterned membrane. Again, a mono-exponential decay transient is observed but with a significantly lower rate of $\tau_1^{-1}$=0.10±0.02ns$^{-1}$. The spectral dependence of $\tau_1^{-1}$ is presented in fig 2b (open circles). A distinct step in $\tau_1^{-1}$ is observed from ~1-0.1ns$^{-1}$ as $\lambda_{det}$ increases from 910-940nm. This corresponds to moving from outside to inside the PBG (c.f. fig 1a) and reflects the reduction of the local density of photonic states.[8] This observation indicates that the edge of the air-band continuum lies close to 910nm for *r/a*=0.3, again in good agreement with the measured cavity mode wavelengths (*M1* and *M2)* and the calculations presented in fig 1a.

Similar time-resolved measurements were recorded from the position of the cavity as a function of $\lambda_{det}$. When $\lambda_{det}$ is spectrally detuned from the cavity mode, we observe a mono-exponential decay transient with a decay rate similar to those recorded on the body of the PC (filled squares – fig 2b). However, as $\lambda_{det}$ approaches the *M1* cavity resonance the decay transient develops a clear bi-exponential character with an upper decay rate faster than our temporal resolution[20] ($\tau_{cav-1}^{-1}$=3-10 ns$^{-1}$) and a lower decay rate comparable to that obtained from position (ii) as discussed above ($\tau_{cav-2}^{-1}$=0.2 ±0.02 ns$^{-1}$). The fast decay originates from QDs which are both *spectrally* and *spatially* coupled to the cavity mode and, therefore, exhibit a large Purcell enhancement of the SE-rate. In contrast, the lower rate stems from QDs that are spectrally in resonance with the cavity mode but weakly coupled since they are spatially displaced from the electric field maximum.[18]

The bottom panel of Fig 2b shows the spectral evolution of the ratio $\tau_0^{-1}/\tau_{cav-2}^{-1}$; a measure of the relative influence of the PBG on the SE-dynamics. For 960<$\lambda_{det}$<1025nm this ratio stays constant at ~10x showing that the SE-rate is suppressed throughout the PBG. As $\lambda_{det}$ reduces towards the air band continuum the ratio approaches unity, showing that $\tau_{cav-2}^{-1}$ tends towards the intrinsic value $\tau_0^{-1}$ for dots in the unpatterned region of the sample.[21]

The modification of the SE rate due to the reduced local density of photon states in the PC is accompanied by spatial redistribution of the emission.[8] In a simplified picture, the total recombination rate ($\Gamma_T$) for QDs in the slab can be written

$$\Gamma_T = \Gamma_{nr} + \Gamma_z + \Gamma_{x,y} \qquad \text{eqn 1}$$

where $\Gamma_{x,y}$ and $\Gamma_z$ denote the emission rates for carriers coupled to the in-plane and vertical emission channels, respectively, and $\Gamma_{nr}$ is the non-radiative recombination rate. In the case of the unpatterned GaAs membrane $\Gamma_z \ll \Gamma_{x,y}$ such that $\Gamma_T$ is dominated by coupling to guided modes in the slab waveguide. When the 2D PC is introduced, in-plane emission is strongly inhibited as discussed above ($\Gamma_{x,y} \sim 0$) and $\Gamma_T = \Gamma_{nr} + \Gamma_z$. In this case, $\Gamma_T$ is dominated by radiation into vertical emission channels despite $\Gamma_z$ being small. In an effort to separate the decay dynamics from spatial redistribution effects, we compared the spatial dependence of the SE-rate with the *time integrated* emission intensity following pulsed excitation. This quantity, labelled $\langle I \rangle_t$ in fig 3, is a direct measure of the photon extraction efficiency ($\xi \sim (\Gamma_z + \Gamma_{nr})/\Gamma_T$). Fig 3 compares the spatial dependence of the SE-rate (lower panel) with $\langle I \rangle_t$ (upper panel) along a cross section of the structure through the cavity for $\lambda_{det}=\lambda_{M1}$ (filled squares) and $\lambda_{det}=\lambda_{M1}+20$nm (filled circles). For QDs in the unpatterned GaAs-AlGaAs waveguide and the underetched GaAs membrane, the SE-rate is mono-exponential with a rate $\tau_0^{-1} \sim 1.25 \pm 0.05$ ns$^{-1}$ as discussed above. We observe a ~3.8x decrease of $\langle I \rangle_t$ as we move from the bulk material onto the unpatterned GaAs membrane. This arises since efficient wave-guiding occurs in the membrane and $\Gamma_z \ll \Gamma_{x,y}$, reducing the fraction of photons leaving the sample vertically. As soon as we probe the dots in the PC, $\Gamma_T$ decreases by $\sim 10 \times$ whilst $\langle I \rangle_t$ simultaneously increases by a similar factor relative to dots in the unprocessed region of the sample. This observation clearly demonstrates that the 2D PC efficiently redistributes the SE energy and, furthermore, indicates that non-radiative recombination is negligible under the experimental conditions studied here. In comparison, the data recorded from the nanocavity show that the SE-rate increases by almost two orders of magnitude when $\lambda_{det}=\lambda_{M1}$ (blue squares) due to the strong Purcell enhanced emission into the cavity mode and efficient collection of the cavity mode radiation by our collection optics. This is accompanied by a ~35 fold increase of $\langle I \rangle_t$ when compared to QDs in the unpatterned region.

To summarize, we conclude that a 2D-PC nanocavity is capable of significantly enhancing both $\Gamma_T$ and the external quantum efficiency. In comparison, the PBG alone reduces $\Gamma_T$ but results in a similar enhancement of $\xi$ compared with unpatterned material. For applications where the external quantum efficiency is most critical, such as single photon generation, the 2D PBG alone may provides a significant practical advantage since it does not require complex spatial positioning of the dot and spectral coupling to a high finesse cavity mode.

**Figure Captions**

FIG. 1. (color online). (a) Plan view SEM image indicating the positions of interest (i)-(iv) as discussed in the text. The main panel shows 3D calculations of the photonic bandstructure for a PC nanocavity with a reduced symmetry defect and r/a=0.28-0.32. Continuum bands are denoted by the gray shaded regions enclosing the PBG and cavity modes M1 and M2 (full blue lines). (b) Typical µPL spectrum recorded from the cavity showing the single dot emission throughout the PBG and cavity modes. The calculated D - field distribution for the two cavity modes are also presented.

FIG. 2. (color online). (a) Typical µPL decay transients recorded from QDs emitting in the unpatterned GaAs membrane, the PC and in the defect nanocavity. (b) Spontaneous emission rates as a function of the wavelength for different measurement positions (i) - (iii). The full gray line shows the µ-PL spectrum for comparison. (lower panel) Ratio of the measured decay rates from QDs in the unpatterned region of the sample and the slow component close to the cavity site.

FIG. 3. (color online). Spatially dependent measurements of SE-dynamics and photon collection efficiency, tracing a line through the centre of the cavity. (lower panel) SE rate for QDs emitting at $\lambda_{det}=\lambda_{M1}$ (filled squares) and $\lambda_{det}=\lambda_{M1}+20nm$ (filled circles), respectively. (upper panel) The corresponding time integrated PL intensity $<I>_t$ for the same two detection wavelengths.

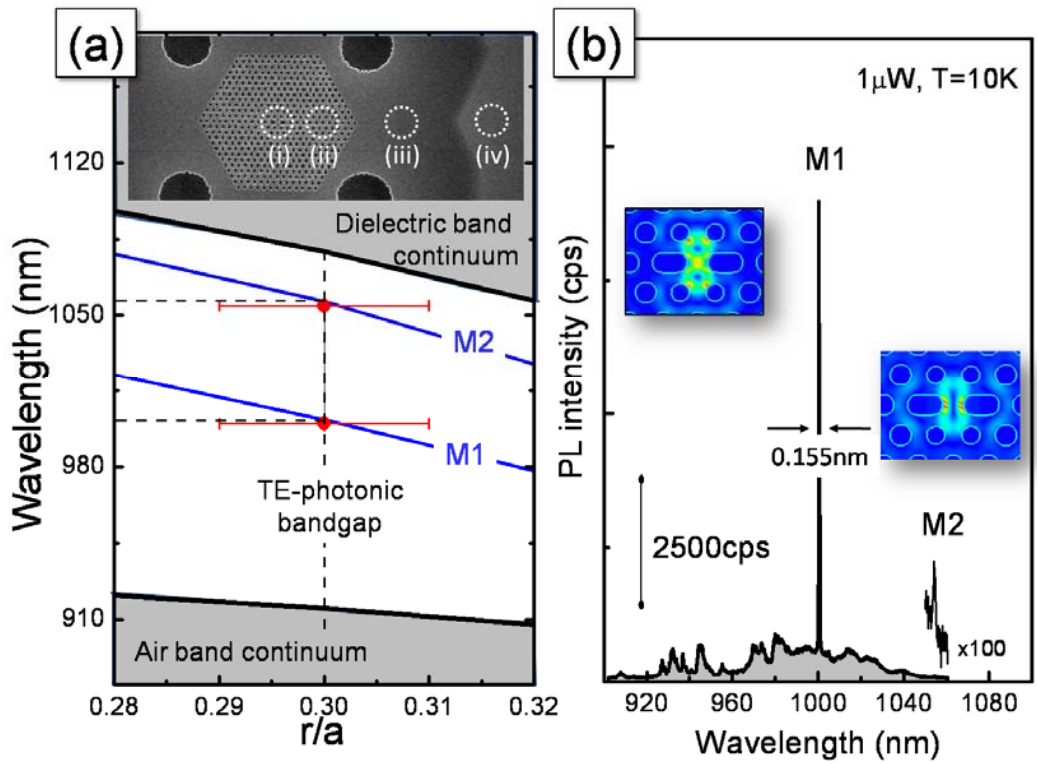

**Figure 1**

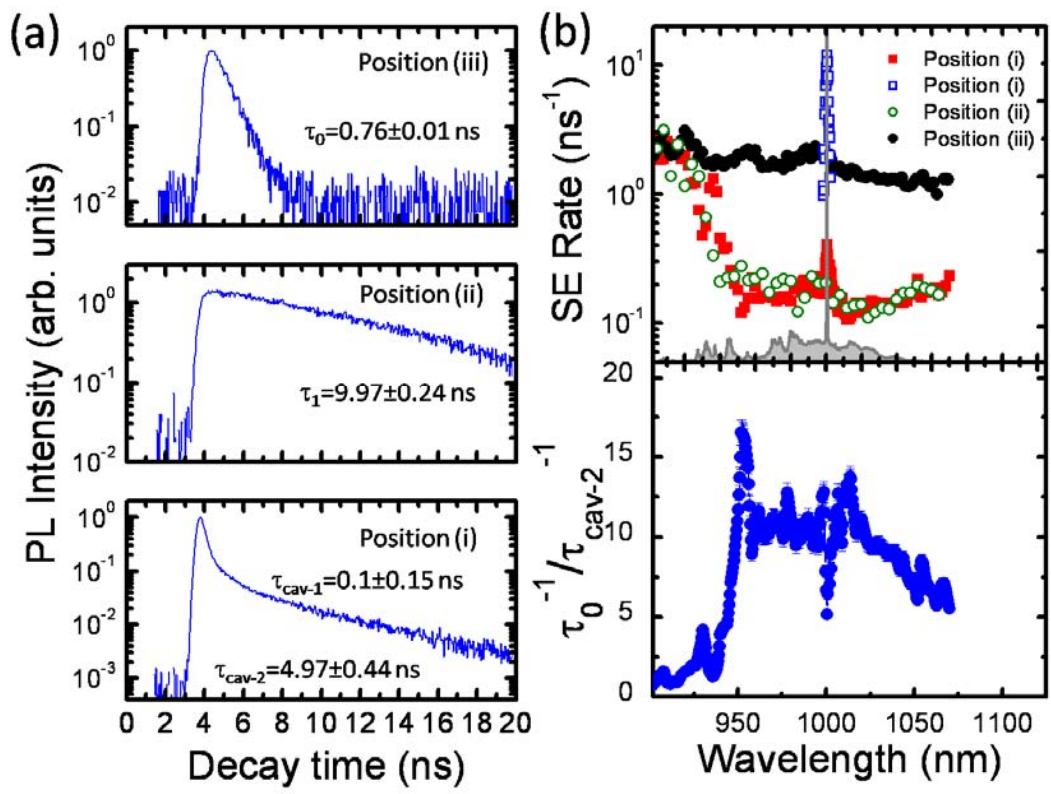

**Figure 2**

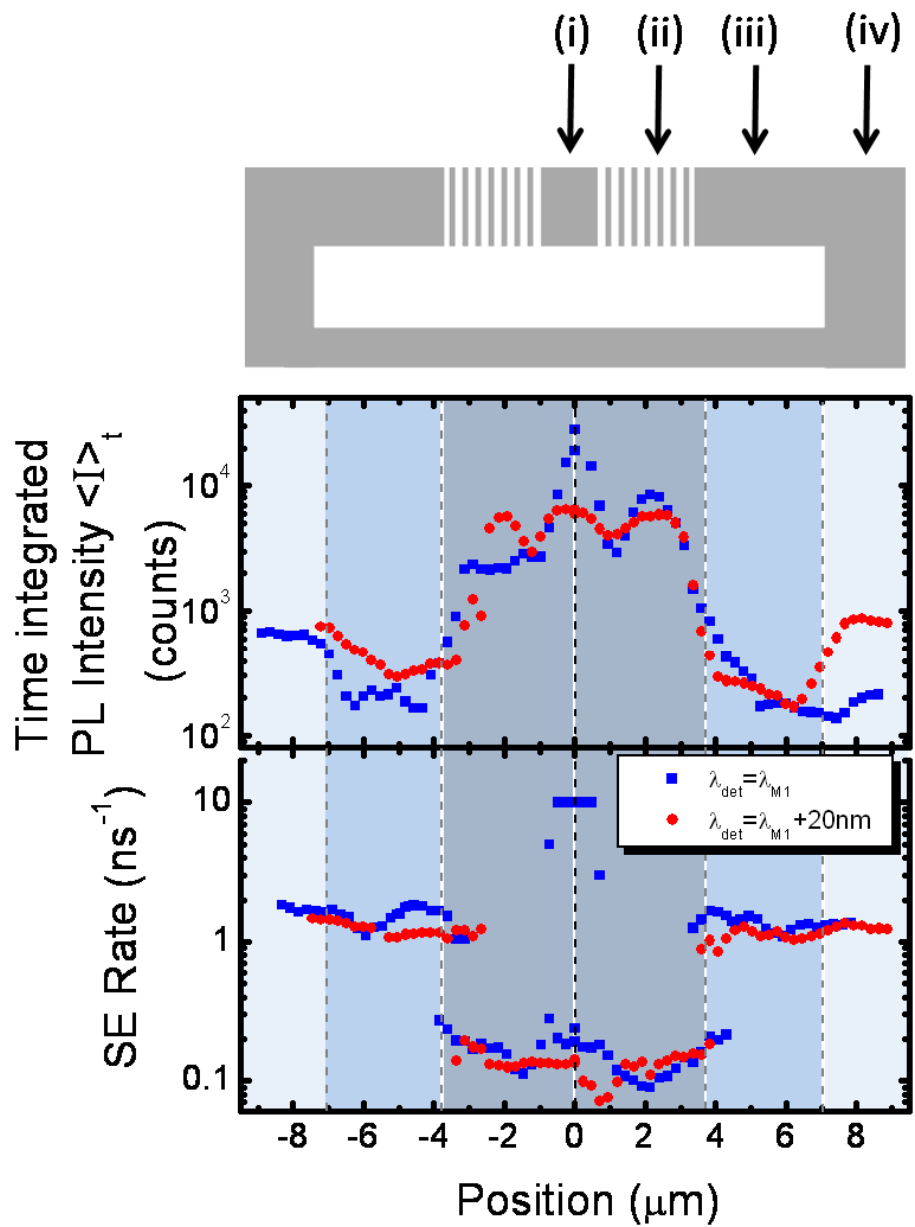

**Figure 3**

[19] Mode M2 appears to be weak in the data presented in fig 1b due to the low sensitivity of the Si-CCD camera for $\lambda > 1\mu m$.
[20] We determined the fast component of the decay transient more accurately using a Streak camera that provided a temporal resolution better than ~20ps. The fast SE-decay rate was found to be $22\pm 8 ns^{-1}$, resulting in an experimentally measured Purcell enhancement of $\tau_0/\tau_2 = 18 \pm 5$.
[21] We are not able to determine the spontaneous emission rate for $\lambda > 1070nm$, since the quantum efficiency of our detector decreases significantly above 1000 nm. Moreover, the QD ground state of our sample emits at ~1020nm.